\documentclass[twoside]{ilcws07}
\usepackage[latin1]{inputenc}
\usepackage[dvips]{graphicx,epsfig,color}
\usepackage{wrapfig,rotating}
\usepackage{amssymb,amsmath,array}

\begin{document}
\title{
%
A GEM TPC End Panel  Pre-Prototype}
%
%
\author{A.Ishikawa$^1$, A.Aoza$^1$, T.Higashi$^1$, A.Sugiyama$^1$, H.Tsuji$^1$,
Y.Kato$^2$, K.Hiramatsu$^2$,\\ T.Yazu$^2$, 
T.Watanabe$^3$, 
O.Nitoh$^4$, H.Ohta$^4$, K.Sakai$^4$, H.Bito$^4$,\\
K.Fujii$^5$, M.Kobayashi$^5$, H.Kuroiwa$^5$, T.Matsuda$^5$, R.Yonamine$^5$,\\
Y.Gao$^6$, Y.Li$^6$, J.Li$^6$, L.Cao$^6$, Z.Yang$^6$,
A.Bacala$^7$, C.J.Gooc$^7$, R.Reserva$^7$, D.Arogoncia$^7$
\vspace{.3cm}\\
1- Saga University, Honjo-machi 1, Saga 840-8502, Japan\\
2- Kinki University, 3-4-1 Kowakae, Higashi-Osaka 577-8502, Japan\\
3- Kogakuin University, 1-24-2 Nishi-Shinjyuku, Tokyo 163-8677, Japan\\
4- Tokyo University of Agricalture and Technology, Koganei, Tokyo, 184-8588, Japan\\
5- IPNS, KEK, Oho 1-1, Tsukuba 350-0810, Japan\\
6- Tsinghua University, CN-100 084, Beijing, China\\
7- Mindanao State University Iligan Institute of Technology, 9200 Lligan, Philippines\\
}

\maketitle

\begin{abstract}
A GEM TPC end panel pre-prototype  was constructed for a large LC-TPC prototype 
to test its basic design philosophy and some of its engineering details.
Its interim test results are presented.

\end{abstract}

\section{Introduction}
Three out of the four detector concepts for the ILC feature 
a Time Projection Chamber~(TPC) as their main trackers.
The LC-TPC is required to have more than 100 sampling points 
with a spatial resolution of 150$\,\mu$m or better 
and a two-track separability down to 2$\,$mm.
%
Beam tests of a small prototype~\cite{Arogancia:2007pt,Dixit:2003qg} 
have shown that
the required  performance is achievable if we adopt a pad width of
as narrow as $\sim$1$\,$mm  
or a resistive anode to spread the signal charge on a pad plane. 
Basic properties of a TPC with a micro pattern gas detector~(MPGD) readout plane 
are also understood through these small prototype tests.
%
Our next target is to construct
a large prototype having multi-MPGD-panels with small readout pads, 
and to demonstrate the required performance
under more realistic experimental conditions with panel boundaries and distortions,
thereby allowing a smooth extrapolation to the real LC-TPC. 
The LCTPC collaboration plans to have a beam test with a large prototype at DESY
in the summer of 2008. 
%
%
For the beam test, we will prepare Gas Electron Multiplier~(GEM) panels.
%
Our basic design goals for the pre-prototype GEM end panel includes the following:
to allow the required smallness and density of readout pads,
to minimize dead spaces due to phi-boundaries of adjacent panels, and
to allow easy replacement of GEM foils when necessary.
Considering these requirements, we decided to 
use as large GEM foils currently available as possible, and
to stretch the GEM foil in the radial directions so as to avoid thick mullions
introducing dead regions for high momentum tracks.
We also designed a GEM mounting system though which  we can
supply necessary high voltages to the GEM foils.
Prior to the production of GEM panels for the beam test,
we constructed a pre-protoype to test its basic design philosophy and
some of its engineering details including fabrication methods for the GEM end panels.

\section{Construction of the GEM TPC End Panel Pre-Prototype.}
\subsection{GEM}
%
Our GEM foils are supplied by Scienergy Co.~\cite{Scienergy} and
consist of 5$\,\mu$m thick copper electrodes sandwiching
a 100$\,\mu$m thick liquid crystal polymer insulator.
%
The thick GEM foils allow stable operation with higher gain
than popular 50$\,\mu$m thick GEM foils. 
A double GEM configuration is hence enough to give a gain of more than $10^4$.
The hole diameter and pitch are 70$\,\mu$m and 140$\,\mu$m, respectively.
The hole shape is cylindrical due to dry etching unlike 
that of CERN GEMs, which is biconial.
The active area is fan-shaped spanning $9.2^\circ$ in the $\phi$ direction with inner 
and outer radii of $128$ and $139\,$cm ($\sim$ $20 \times 11\,$cm${}^2$),
which is about 2 times larger than the GEM foils we used for our small prototype. 
To keep the stored energy small enough,
the GEM electrodes are divided into two in the radial direction with a boundary of 100$\,\mu$m.

A set of G10 frames should be glued to each GEM foil in order to be mounted on a readout PC board. 
For the frame gluing we developed a GEM stretcher, which consists of two parts.
One is an acrylic GEM stretcher frame and the other is a set of a middle frame and a GEM adjuster
made of aluminum. 
The lower part of the acrylic stretcher frame has a groove with a depth of 2$\,$mm 
and the upper part has screw holes aligned to it.
The middle aluminum frame has the same size as the groove.
By sandwiching a GEM foil with the lower and the upper parts of the acrylic stretcher frame 
together with the aluminum frame, 
and by screwing bolts into the holes of the upper piece and pressing down the aluminum frame into the 
groove, we could stretch the GEM foil.
The GEM foil, the G10 frames, and the adjuster have through holes aligned to each other. 
We stacked them up together, put pins into the holes of the adjuster to align them,
and glued the G10 frames with epoxy adhesive to the GEM foil.
Notice that the frames covered only the inner and outer  edges of the GEM foil to reduce the dead space pointing 
to the interaction point. 
This fabrication method has been established and allowed us to produce 3 panels per day.

\subsection{Readout PC Board}
The electrons multiplied by the GEM foils are read out by a PC board.
The PC board is $2\,$mm${}^t$ thick and has a size enough to cover the GEM foils, spanning
$9.2^\circ$ in $\phi$ and having inner and outer radii of $127$ and $140\,$cm, which are
about $1\,$cm extended in both inner and outer directions to facilitate the GEM mounting.
It carries 20 pad rows on its front side of which the inner 10 have 176 pads each
and the outer 10 have 192 pads each,
with every two rows staggered by half a pad width to minimize the so-called hodoscope effect.
A typical pad size is about $1.1\times5.5\,$mm which is small enough for the intrinsic charge spread of $\sim350\,\mu$m.
The pads are wired to readout connectors on the back side of the PC board through a five-layer FR4.
The PC board, being of six layers, 
has no through-holes due to wiring,
thereby assuring the gas-tightness required for the TPC operation.

We used connectors supplied by JAE~\cite{JAE}. 
Each connector has 40 channels, of which 32 are used for signal readout and the remaining 8 for ground.
Its size is about $15\times5\,$mm$^2$, which is one of the smallest connectors commercially available. 
Pre-amplifiers are connected to the PC board with flexible cables. 
For the large prototype, pre-amplifiers and flash ADCs will be mounted on 
small PC boards and are directly connected to the readout PC board. 
We will not use the connectors for the real ILC TPC,
since pre-amplifiers and flash ADCs will be mounted on the surface of the PC board.


High voltage~(HV) electrodes are also wired through the PC board.
To apply the HVs to the GEM electrodes, we adopt a bolt-and-nut method. A brass nut is adhered to an electrode on
the front side of the PC board. 
We tried two methods to fix the nut. One is soldering, the other is gluing with a conductive paste 
(dotite by Fujikura Kasei Co.~\cite{FK}). It turned out that soldering is much stronger than the conductive paste,
besides the dotite produces threads. We will hence use soldering for the large prototype construction.
%
The lower and the upper GEM foils are stacked on the PC board, and bolted through their G10 frames
to bite the GEM electrodes so as to supply required HVs through the bolts.
The spacings between the GEM foils and the readout pad plane are determined by the thicknesses of the
G10 frames, resulting in a transfer gap of 4$\,$mm and an induction gap of 2$\,$mm.
An aluminum flange with a groove for an O-ring is glued to the back side of the PC board with epoxy adhesive 
to avoid mechanical distortion and to be mounted to a gas container. When we glued the aluminum flange, 
we aligned the flange and the PC board by hand since there are no alignment 
holes or posts in the PC board and the flange. 
The positions of the readout pads are determined by this flange, so some alignment
posts should be prepared for the large prototype. 
The GEM end panel was mounted on a gas container with 16 M3 bolts.

\section{Measurement of Gain Uniformity over the Panel.}
The pre-prototype chamber is filled with a 90:10 mixture of Ar and iso-butane gases.
We applied 410$\,$V to each of the two GEM foils, and electric fields of 100, 2050, and 3075$\,$V/cm 
to the 25$\,$mm long drift, the 4$\,$mm long transfer, and the 2$\,$mm long induction regions, respectively. 
Under these conditions, the gas gain was about $2 \times 10^4$  and the signal spread was about $550\,\mu$m. 
We irradiated the pre-prototype panel with X-rays from ${}^{55}$Fe through the windows of the test chamber
to measure the gain uniformity over the panel.

\begin{wrapfigure}{r}{0.5\columnwidth}
\centerline{
\includegraphics[width=0.45\columnwidth]{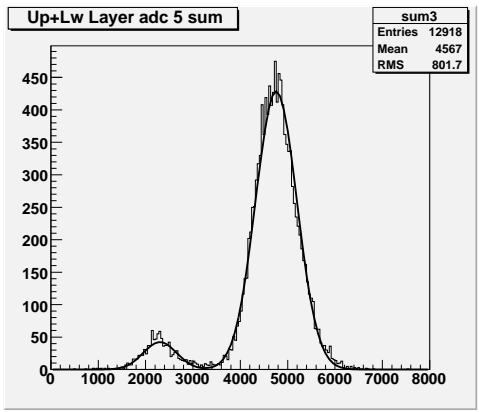}
}
\caption{Charge sum distribution.}\label{Fig:ADC}
\end{wrapfigure}

First, we checked the charge distribution over the readout pads. 
Since the signals sometimes spread over 2 pad rows,
we required the signal charge be shared by 2 pad rows
and summed  the signals over 5 contiguous pads on each of these  two rows to avoid mis-collection.
Figure$\,$\ref{Fig:ADC} shows the charge sum distributions for 10 pads.
Both a 5.9$\,$keV main peak and a 2.9~keV escape peak can be clearly seen.

\begin{wrapfigure}{r}{0.5\columnwidth}
\centerline{
\includegraphics[width=0.45\columnwidth]{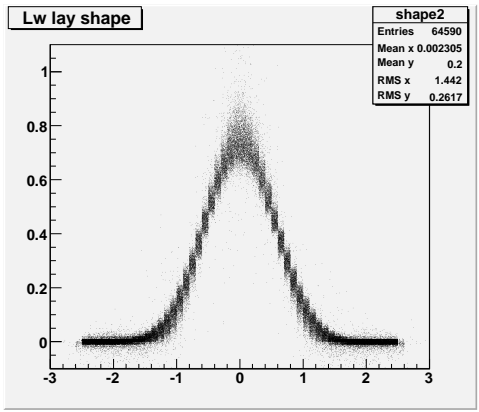}
}
\caption{Signal spread.}\label{Fig:signal}
\end{wrapfigure}

Second, we checked the charge spread. A center of gravity was calculated from the measured pad signal charges
and their $\phi$ positions as well as
the charge fraction on each pad.
By plotting the charge fraction as a function of the charge center measured from the middle of the central pad,
we could get an image of the charge distribution over the pad plane.
Gaussian fit to the distribution resulted in
a (1-$\sigma$) width of about 550$\,\mu$m corresponding to half a pad width
as expected (Fig.~\ref{Fig:signal}) from the diffusion.

We then measured the charge sum distributions at 28 positions over the panel for the uniformity test,
usually requiring the charge sharing.
We found, however, that charge sharing never happened near the boundary of the inner and the outer GEM electrodes.
The exact reason is still unknown but it could be attributed to the charge-up of the insulator of the GEM boundary
affecting the charge collection.
The charge sharing was hence not required for the positions in the boundary region. 
%
After normalizing the charge sums to that of some reference position,
we found that the normalized charge sums range from 0.49 to 1.08.
%
%
The observed gain non-uniformity is 2.5 times larger than the expected 20\% or less 
from the mechanical tolerance of the panel.
%
We  found a large field distortion near the panel edges, which partly explains
the non-uniformity but not all of it.
Further investigations for possible causes are needed including variations of 
operation conditions such as gas concentration, etc..

\section{Summary}
We have constructed and tested a pre-prototype of the GEM TPC end panels to 
verify basic design philosophy and some of engineering details including fabrication methods
for the large prototype of the real ILC TPC.
We have basically established a GEM framing scheme with some
minor problems to be improved for the large prototype construction. 
We have also measured the gain uniformity over the pre-prototype panel
and observed a 50\% non-uniformity at maximum.
The non-uniformity could partly be attributed to the field distortion due to the test chamber setup,
but requires further studies to fully validate our basic design philosophy.

\section*{Acknowledgements}
This study is supported in part by the Creative Scientific Research 
Grant No. 18GS0202 of the Japan Society for Promotion of Science.

\begin{footnotesize}



%

\end{footnotesize}


\end{document}